# Spontaneous Exact Spin-Wave Fractals in Magnonic Crystals

Daniel Richardson[1], Boris A. Kalinikos[1,2], Lincoln D. Carr[3], and Mingzhong Wu[1†]

[1]Department of Physics, Colorado State University, Fort Collins, CO 80523, USA
[2]St. Petersburg Electrotechnical University, 197376 St. Petersburg, Russia
[3]Department of Physics, Colorado School of Mines, Golden, CO 80401, USA

Exact fractals of nonlinear waves that rely on strong dispersion and nonlinearity and arise spontaneously out of magnetic media were observed for the first time. The experiments make use of a microwave to excite a spin wave in a quasi-one-dimensional magnonic crystal. When the power of the input microwave ($P_{in}$) is low, the output signal has a power-frequency spectrum that consists of a single peak. When $P_{in}$ is increased to a certain level, new side modes are generated through modulational instability (MI), resulting in a comb-like frequency spectrum. With a further increase in $P_{in}$, each peak in the frequency comb can evolve into its own, finer comb through the MI. As $P_{in}$ is increased further, one can observe yet another set of finer frequency combs. Such a frequency-domain fractal manifests itself as multiple layers of amplitude modulation in the time-domain signal.

A fractal is a shape made of parts each of which is similar to the whole in some way. One can group fractals into two main categories, (i) exact fractals (or regular fractals) in which the same feature replicates itself on successively smaller scales and (ii) statistical fractals (or random fractals) that display statistically similar features.[1,2,3] Statistical fractals have been observed in a rather wide variety of physical systems, ranging from material structures to lungs in human bodies and stock price fluctuations. In stark contrast, exact fractals are relatively rare in nature, though they can be very easily constructed by mathematical models. Examples of exact fractals include optical fractals formed using self-similar structures.[4,5]

Despite the above facts, exact fractals have been found in nonlinear dynamics, which is rather surprising in view of the strong sensitivity of nonlinear systems. They are space-domain soliton fractals, demonstrated numerically, and time-domain soliton fractals, observed experimentally. The realization of the first one relied on the use of a one-dimensional (1D) nonlinear waveguide that consists of different sections, each with a larger dispersion coefficient $D$ than the prior section.[6,7] As a soliton in the first section enters the next section, it experiences an abrupt increase in $D$ and thereby breaks up into several smaller solitons or daughter solitons. When the daughter solitons enter the next section, each of them undergoes another breakup and produces even smaller solitons or granddaughter solitons. Thus, successive changes in $D$ create soliton fractals along the waveguide.

The demonstration of the time-domain soliton fractals made use of a feedback ring that consisted of a 1D nonlinear media, and an amplifier that amplified the output signal from the media and then fed it back to the input of the media.[8] With an appropriate amplification, a single soliton is self-generated in the ring; as the soliton circulates in the ring, its amplitude varies in a fractal manner, yielding a time-domain fractal. In this case, the amplifier ensures sufficient nonlinearity needed to maintain the soliton, while the periodic feedback modifies the wave dispersion to enable the fractal dynamics.

This letters report on the observation of a new type of exact fractals in nonlinear dynamics that, in contrast to the space- and time-domain soliton fractals, form spontaneously out of the constituent media, without being forced into being, and do not involve solitons. To make an analogy, there is a strong difference between spontaneous symmetry breaking, integral to the theory of phase transitions, and forcing symmetry-breaking. The observation uses nonlinear media in which a spatial periodic potential is introduced to create strong dispersion that facilitates the formation of a fractal. The experiments utilize a quasi-1D magnonic crystal[9,10,11,12,13,14,15,16,17] that consists of a long and narrow magnetic $Y_3Fe_5O_{12}$ (YIG) film strip with periodic transversal lines etched into the film. This medium supports the propagation of spin waves. The etched lines create a periodic potential for spin waves, and the latter leads to significant modification to the spin-wave dispersion curve at certain wavenumbers associated with the dimensions of the periodic lines.[10,11] Upon the excitation of a continuous spin wave in one end of the media, the spin wave propagates to the other end, resulting in an output signal that manifests itself as a single peak in the power-frequency spectrum. With an increase in the input power $P_{in}$, the initial peak (or the mother) can produce additional side peaks (or the daughters) in the frequency range with strong dispersion through modulational instability (MI),[1,18,19,20] resulting in a comb-like frequency spectrum. As $P_{in}$ is further increased, each peak in the comb evolves into its own, finer frequency comb (granddaughters), also through the MI. Such frequency fractals, which are illustrated in Fig. 1, manifest themselves as multiple layers of amplitude modulation in the time-domain signal.

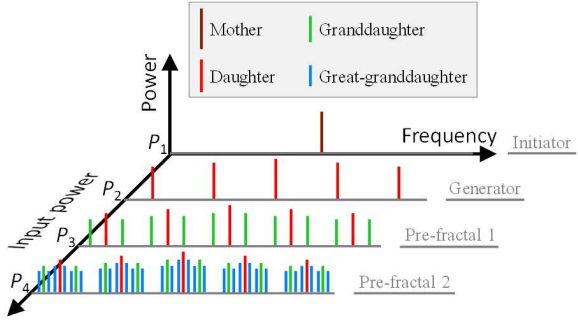

Fig. 1. Illustration of the development of a frequency fractal with an increase in the input power ($P_1<P_2<P_3<P_4$).

Three important points should be highlighted. First, the new fractals in this work are fundamentally different from the previously demonstrated soliton fractals. On the one hand, those fractals are for solitons, which involve a fine balance between the dispersion-induced pulse broadening and nonlinearity-caused self-narrowing,[6-8] while the new fractals do not require such a balance. In this aspect, this work indicates that exact fractals in nonlinear systems do not have to involve solitons. On the other hand, the new fractal relies on a completely different approach to realize the conditions needed for fractal formation; it makes use of spatial periodic potentials to achieve strong dispersion required by fractal generation, while those soliton fractals use successively increased dispersion and periodic feedback, respectively, to interrupt soliton dynamics and realize fractals. Thus the new fractals are spontaneous, not forced. Second, the approach in this work is of a general nature and can be applied to achieve similar fractals in other nonlinear systems, including electromagnetic transmission lines, optical fibers, and water waves. Finally, in addition to advancing the field of fractals, the results also help interpret various nonlinear effects in magnonic crystals, such as instability and nonlinear damping.[15,17]

The experimental configuration is sketched in Fig. 2(a). The experiments made use of a quasi-1D magnonic crystal that consisted of a 10-mm-long, 2.5-mm-wide, 10.3-μm-thick YIG film strip with 12 lines etched into the film. Each etched line is 50 μm wide and 3.3 μm deep, and the spacing between the lines is 400 μm. Backward volume spin waves[21,22] are excited by placing a microstrip line on one end of the YIG strip and feeding it with microwaves, and are detected by a second microstrip line placed on the other end of the YIG strip.

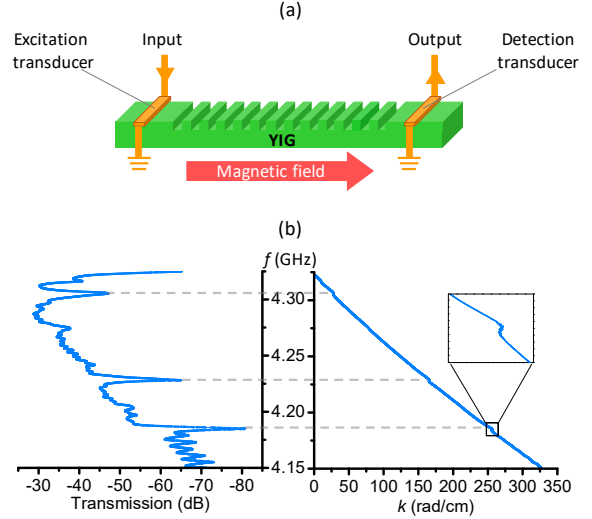

Fig. 2. (a) Experimental configuration. (b) Transmission profile (left) and frequency $f$ vs. wavenumber $k$ dispersion curve (right) measured with a YIG-based 1D magnonic crystal for $P_{in}$=0.7 mW.

The separation of the two microstrip lines are about 5.5 mm. The magnetic field, indicated by the red arrow, is kept constant at 1175 Oe. Figure 2(b) shows the transmission profile and frequency ($f$) vs. wavenumber ($k$) dispersion curve obtained from the transmission coefficient measurements on the magnonic crystal. The strong dips in the transmission profile and the corresponding jumps in the dispersion curve represent unique characteristics of the magnonic crystal[10,11,17] that result from the periodically etched lines.

The data in Fig. 2(b) were measured with a relatively low input power ($P_{in}$=0.7 mW) over a relatively wide $f$ range using a vector network analyzer. In contrast, Fig. 3 presents the data measured at a significantly higher power ($P_{in}$=7 W) over a much narrower $f$ range which are both relevant to the fractal measurements described later. The dispersion data, shown by the dots in Figs. 3(c) and 3(d), were interpolated to produce dispersion curves, shown by lines, and the latter were used to numerically determine the dispersion coefficient $D = \frac{d^2(2\pi f)}{dk^2}$ presented in Figs. 3(e) and 3(f). The data in Fig. 3 clearly show that, as one sweeps $f$ across a transmission dip, the dispersion coefficient $D$ can become substantially large and can even flip its sign. To be more specific, |$D$| is about $10^3$ cm$^2$/(rad·s) in the off-dip region, which is close to typical values in continuous YIG thin films,[22] but can be seven orders of magnitude larger in the transmission dip. It is this strong dispersion that enables the formation of the fractals presented below.

Figure 4 shows four power-frequency spectra measured at different $P_{in}$, as indicated, using a spectrum analyzer that demonstrate the development of the spin-wave fractal. At $P_{in}$=0.7 mW, the spectrum consists of only one peak, as shown in Fig. 4(a), at the frequency that is exactly equal to the input frequency. This peak corresponds to the initiator or the mother shown in Fig. 1. As $P_{in}$ is increased to 7 W, several new side peaks are generated through the MI,[1,18-20] and the initial single-peak spectrum evolves into a frequency comb, as shown in Fig. 4(b). The comb spectrum corresponds to the generator in Fig. 1, and the new peaks can be termed as daughter modes. With a further increase in $P_{in}$, each peak in the comb develops its own, finer comb (granddaughters), as shown in Fig. 4(c), also through the MI. As $P_{in}$ is increased further, each granddaughter generates several great-granddaughters, as shown in Fig. 4(d). The spectra in Figs. 4(c) and 4(d) correspond to pre-fractal 1 and pre-fractal 2, respectively.

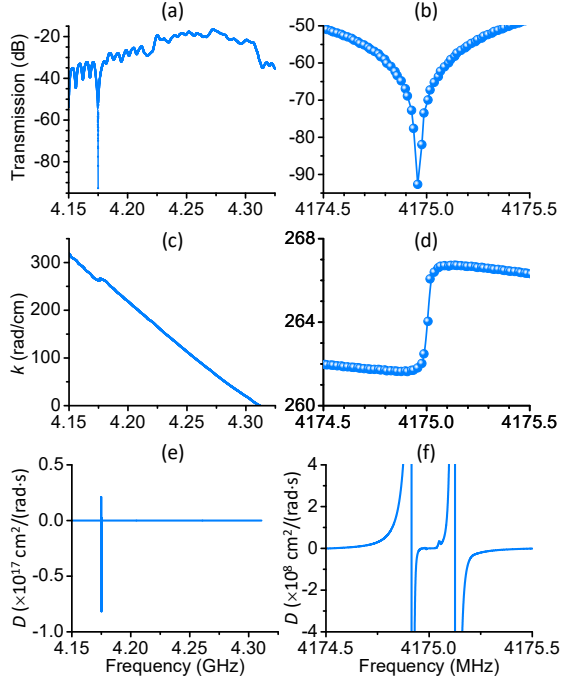

Fig. 3. The first and second rows show the transmission profile and the frequency vs. wavenumber ($k$) dispersion curve, respectively, measured with a YIG-based 1D magnonic crystal for $P_{in}$=7 W. The third row shows the dispersion coefficients ($D$) calculated based on the dispersion curves in the second row. The right column shows the same data as in the left column but over a much narrower frequency range.

The time-domain signals, measured by a fast oscilloscope, that correspond to the spectra in Figs. 4(a), 4(b), and 4(c) are presented in Figs. 5(a), 5(b), and 5(c), respectively. One can see that, with an increase in $P_{in}$ from 0.7 mW to 7 W, the originally constant envelope of the time-domain signal breaks up into a periodic modulation with a period of ~2.0 µs which is the exact reciprocal of the frequency spacing of the comb spectrum in Fig. 4(b). As $P_{in}$ is further increased to 14 W, a secondary modulation with a much longer period appears on the top of the first modulation. The period of this modulation is ~20 µs which is the reciprocal of the spacing of the daughter combs in Fig. 4(c).

The physical process that underlies the above-mentioned MI is four-wave interaction;[1,19] and in magnetic materials such an interaction is often termed as four-magnon scattering.[23,24] The process satisfies the energy conservation law $2\omega_0 = \omega_1+\omega_2$, where $\omega_0$ and $\omega_{1,2}$ are the frequencies of the initial mode and the new side modes, respectively. When sufficiently strong, the side mode ($\omega_1$ or $\omega_2$) can interact with the initial mode ($\omega_0$) to produce additional side modes through the four-wave interaction, resulting in an overall comb-like spectrum. As the conservation law can be rewritten as $\omega_2 - \omega_0 = \omega_0 - \omega_1$, one can expect the formation of a uniform comb that has an equal frequency spacing $\Delta\omega$. Indeed, all the combs in Fig. 4 are equally spaced. The spacing $\Delta\omega$ generally scales with $\frac{1}{\sqrt{|D|}}$,[1] and this is why the fractal appears in the transmission dip region only. In other words, the significantly enhanced dispersion in the transmission dip (see Figs. 3(e) and 3(f)) enables the generation of the modes with small $\Delta\omega$ and thereby facilitates the formation of new, finer frequency combs. The MI rate, namely, the rate of the instability growth,[1] generally scales with the square of the wave amplitude, $|u|^2$. Since the spin-wave amplitude increases with $P_{in}$, this explains why the fractal develops or evolves to higher levels only at high $P_{in}$.

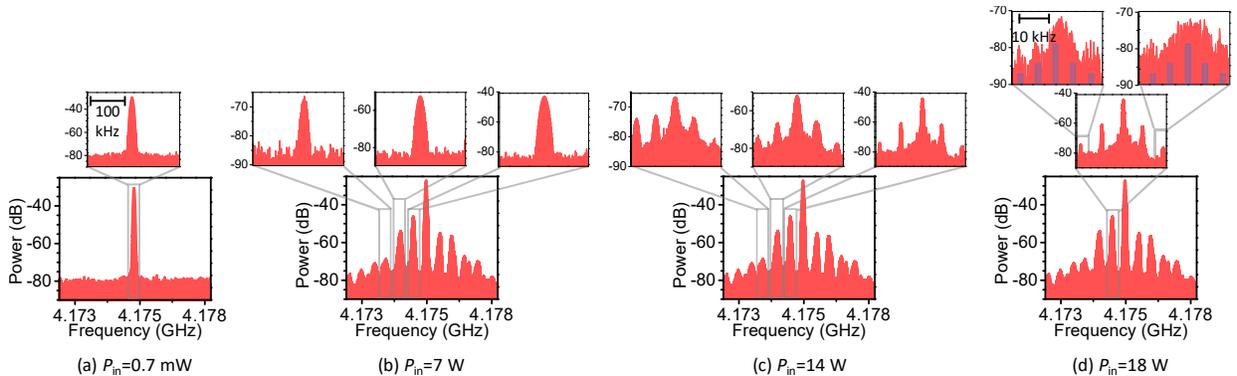

Fig. 4. Power-frequency spectra measured at different $P_{in}$, as indicated, demonstrating fractal development. The diagrams in the middle and top rows share the same frequency scale indicated in the left-most diagram in each row. The vertical bars in the top-row diagrams serve as visual guides to indicate the positions of the frequency peaks.

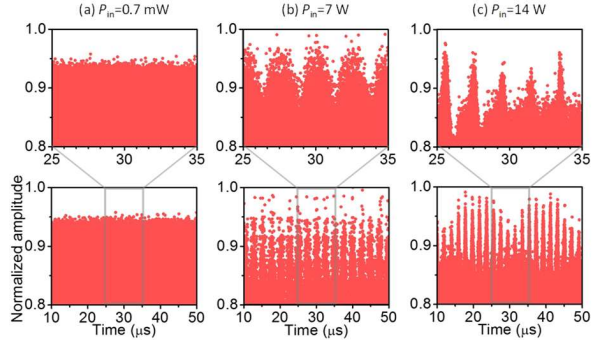

Fig. 5. Time-domain signals measured at different $P_{in}$, as indicated, demonstrating the development of two layers of MI.

Several notes should be made about the fractal data shown in Figs. 4 and 5. First, the granddaughter and great-granddaughter combs are presented only for three selected daughter modes in Fig. 4, but they in fact also exist for other daughter modes. Second, the mother mode has less-developed granddaughter and great-granddaughter combs than the daughter modes. This is probably because the frequency of the mother mode is closer to the center of the transmission dip where $D$ may have a relatively small value as shown in Fig. 3(f). Third, the $\frac{\Delta\omega}{2\pi}$ values for the main combs in Figs. 4(b), 4(c), and 4(d) are 510 kHz, 500 kHz, and 495 kHz, respectively. The slight decrease of $\frac{\Delta\omega}{2\pi}$ with increasing $P_{in}$ is consistent with the facts that one usually has $\Delta\omega \propto |u|$,[1] while the peak intensity of the mother mode slightly decreases with increasing $P_{in}$ because of the re-distribution of energy to new side modes. Fourth, the great-granddaughter modes in Fig. 4(d) would be more visible if the diagrams are enlarged.

Finally, it should be noted that no fractals beyond pre-fractal 2 were observed in the frequency domain and only 2 layers of modulation were measured in the time domain. The main reasons for this include instrumentational limitation (limited sensitivities of the spectrum analyzer and the oscilloscope) and limited nonlinearity due to various sources of damping (for example, two-magnon scattering) and thermal issues arising at very high $P_{in}$. Future work that uses better instrumentation and stronger nonlinearity to demonstrate fractals of more layers is of great interest. It should also be noted that this work involves a constant field along the sample length direction, and a change in the field direction may lead to the absence of the above-presented fractals because both the dispersion and nonlinearity properties of spin waves strongly depend on the equilibrium direction of the magnetization.

In summary, this work demonstrates experimentally the development of an exact fractal for nonlinear spin waves in a quasi-1D magnonic crystal. The fractal exists in the frequency regions where the dispersion is significantly enhanced due to a spatially periodic potential, and is generated through the four-wave interaction. The fractal manifests itself as three layers of comb-like spectra in the frequency domain and two layers of amplitude modulation in the time domain. The new fractal fundamentally differs from the fractals found previously in nonlinear systems; it arises spontaneously out of the fundamental magnonic crystal media, in contrast to previous approaches based on successive forcing of emergent structures, namely, solitons.

This material is based in part upon work supported by the U.S. National Science Foundation under Grants No. DMR-1407962 and No. EFMA-1641989 and the SHINES, an Energy Frontier Research Center funded by the U.S. Department of Energy, Office of Science, Basic Energy Sciences under Award SC0012670. In addition, L.D.C. was also supported by the National Science Foundation under Grants No. PHY-1520915 and No. OAC-1740130 and the U.S. Air Force Office of Scientific Research under Grant No. FA9550-14-1-0287.

†Corresponding author. E-mail: mwu@colostate.edu